# A Wireless Volumetric Metamaterial Resonator for Breast MRI

Yuhan Liu[1,2], Xia Zhu[1,2], Christina A. LeBedis[3], Stephan W. Anderson[2,3*], and Xin Zhang[1,2,4,5,6*]

[1]Department of Mechanical Engineering, Boston University, Boston, Massachusetts, USA
[2]Photonics Center, Boston University, Boston, Massachusetts, USA
[3]Chobanian & Avedisian School of Medicine, Boston University Medical Campus, Boston, Massachusetts, USA
[4]Department of Electrical & Computer Engineering, Boston University, Boston, Massachusetts, USA
[5]Department of Biomedical Engineering, Boston University, Boston, Massachusetts, USA
[6]Division of Materials Science & Engineering, Boston University, Boston, Massachusetts, USA

*Corresponding author: Xin Zhang (xinz@bu.edu); Stephan W. Anderson (sande@bu.edu);
Contributing authors: yuhanyh@bu.edu; xiaz@bu.edu; clebedis@bu.edu

## Abstract

High-performance breast magnetic resonance imaging (MRI) remains limited by the availability, geometry, and cost of dedicated receive coils. In contrast, spine coils are widely installed in MR scanners globally but provide insufficient sensitivity for anterior breast tissue. Here, we report a wireless volumetric metamaterial resonator (VMR) that converts a standard built-in spine coil into a scalable breast MRI receiver platform without modifying the scanner hardware. The VMR consists of axially stacked coaxial resonant rings that support a co-rotating collective mode, producing a centrally concentrated RF magnetic field within the enclosed imaging volume. Leveraging distributed capacitance in the coaxial cable, gap-engineered geometric tuning, and PIN-diode-based self-detuning, the VMR enables size scalability, electric-field confinement, and transmit–receive compatibility, as validated through bench measurements and 3.0 T MRI experiments. Three VMR prototypes spanning small, medium, and large breast-equivalent volumes were designed and evaluated using electromagnetic simulations, bench measurements, and phantom MRI experiments. Compared with a commercial 16-channel breast coil, the VMR-augmented spine coil achieved more than 4-fold higher central SNR and reduced spatial non-uniformity from approximately 70% to 20% within the target imaging region. Under SENSE parallel imaging at acceleration factors of 2-4, the VMR-augmented spine coil maintained an average 4-fold higher central SNR than the commercial coil for three sizes. These results establish the VMR as a passive, scalable hardware strategy for improving breast MRI performance using existing clinical scanner infrastructure.

## Introduction

Breast cancer is the most prevalent malignancy among women worldwide, and its global burden continues to grow: projections estimate 3.2 million new cases and 1.1 million deaths annually by 2050 [1,2]. Early detection remains the most effective strategy available—raising the five-year survival rate from approximately 25% to 99% [3]—yet access to high-quality screening remains profoundly inequitable across regions and healthcare settings [4-7]. This disparity is driven in large part by the unequal distribution of advanced imaging infrastructure, a bottleneck that disproportionately affects low- and middle-income regions but is also present in many under-resourced facilities within developed healthcare systems. Addressing this gap requires not only clinical investment, but hardware innovation that is practical and deployable within the constraints of existing medical infrastructure.

Magnetic resonance imaging (MRI) is widely regarded as the most sensitive modality for breast cancer detection, offering unparalleled soft-tissue contrast, inherently three-dimensional volumetric coverage, and non-ionizing radiation, making it particularly valuable for screening women with dense breast tissue or elevated genetic risk [8-10]. However, the performance of breast MRI is fundamentally constrained by a structural limitation in current hardware architecture. Conventional clinical MRI systems (1.5-3.0 T) rely on large, body-sized birdcage coils for radiofrequency (RF) transmission, paired with surface receive coils to improve local signal [11-14]. While adequate for general imaging, this architecture struggles to provide a high signal-to-noise ratio (SNR), homogeneity, and precise localization for relatively small targets such as the breast [15,16]. These are not limitations that can be resolved through parameter tuning alone, but are rooted in a geometric and electromagnetic mismatch that is inherent to the architecture [17].

Dedicated commercial breast coils—typically incorporating 16 or more receive channels in an organ-conformal geometry—substantially address this mismatch and represent the current standard for high-performance breast MRI [18-21]. However, these systems remain largely confined to well-resourced clinical centers and are unavailable across much of the global MRI-capable infrastructure. By comparison, spine coils are built-in, multi-channel posterior receive arrays integrated into nearly all clinical MRI scanners. Because they are already installed in the patient table and require no additional coil placement, cable connection, or scanner-side hardware setup, repurposing this existing receive infrastructure for breast imaging would substantially reduce the hardware burden, setup complexity, and cost associated with dedicated breast MRI. However, spine coils were designed for posterior anatomical coverage, and their RF sensitivity diminishes rapidly as it extends into the volumetric anatomy of the breast, resulting in insufficient SNR, signal non-uniformity, and compromised diagnostic quality [14-15]. Transforming this widely available but anatomically mismatched hardware into an effective breast imaging receiver therefore represents a clinically meaningful and practically impactful objective.

Recent developments in ultra-high-field (7.0 T) and ultra-low-field ($\leq$0.5 T) MRI have demonstrated promising opportunities to extend MR imaging capabilities beyond conventional field strengths [22,23]. However, current implementations remain largely focused on brain and extremity imaging, with limited demonstrated applicability to breast MRI. More broadly, both directions involve fundamentally new hardware and software ecosystems that are incompatible with the existing installed base of clinical scanners, restricting their deployment to highly specialized scenarios and limiting near-term clinical translation. These considerations point toward a complementary strategy: rather than replacing existing infrastructure, developing solutions that augment breast imaging performance within the widely deployed field strength (1.5-3.0 T) clinical systems, without requiring modifications to scanner hardware or acquisition workflows.

Wireless metamaterial resonators have emerged as a practical approach for this type of augmentation [24-36]. They operate through wireless inductive coupling to the birdcage coil, eliminating the need for excessive cabling, electronic panels, interfaces, adapters, or baluns. Implemented as passive, receive-mode structures that require no direct electrical connection to the scanner, these devices can be designed to conform to diverse anatomical geometries, operate across different scanner platforms, and integrate into clinical workflows without modifications to pulse sequences or system software. Through resonantly enhanced near-field coupling, they reshape the local RF magnetic field distribution and improve SNR in targeted anatomical regions. Despite these advantages, most reported metamaterial resonator designs exhibit limitations that have restricted their broader utility. Reported SNR gains, while measurable, have often been modest in magnitude. More importantly, most designs have not been optimized or evaluated for parallel imaging—a technique that enables scan time reduction through simultaneous multi-channel spatial encoding and is routinely employed in clinical breast MRI—limiting their practical applicability under accelerated imaging conditions [37-40]. A more fundamental constraint lies in the geometry of most existing designs: the majority adopt planar configurations, which, much like a surface coil, provide strong sensitivity near the device surface but suffer from inherently limited penetration depth and rapid signal fall-off into deeper tissue [41-45]. This intrinsic limitation compromises both SNR and field homogeneity across volumetric anatomical targets—a shortcoming that is particularly consequential for breast imaging, where sensitivity throughout the full tissue depth is diagnostically essential. Transitioning from planar to volumetric metamaterial geometries, however, introduces substantially greater design and experimental complexity and additionally requires that the electromagnetic field orientation—including flux linkage and induced current distribution—be carefully matched to the target anatomy and receive coil configuration.

A small number of designs have been developed with precisely this intent, targeting breast MRI specifically, including ceramic resonators [46], circularly polarized metamaterials [47], and quadrature-mode coils [48,49], each of which has demonstrated measurable SNR improvements. However, several important shortcomings persist across these approaches. Many designs lack intrinsic self-detuning mechanisms, requiring careful external management during the RF transmit phase to avoid unintended energy absorption and mitigate RF heating risks—a practical concern that complicates safe clinical integration [32-34,50,51]. Furthermore, the majority have been demonstrated in conjunction with volume transmit coils such as birdcage coils, rather than the spine coil configuration, this architectural choice also limits compatibility with parallel imaging frameworks [37-40]. In addition, existing designs have generally been validated for a single fixed geometry, without demonstrated adaptability to accommodate the range of breast sizes and anatomical variability encountered in clinical practice. As a result, the field currently lacks a metamaterial-based solution that simultaneously delivers substantial SNR improvement, supports parallel imaging, accommodates anatomical variability, and is validated within a clinically realistic spine coil context.

Here we present a modular, geometrically scalable wireless volumetric metamaterial resonator (VMR) that transforms the built-in multi-channel spine coil into a high-performance breast MRI receiver platform. The framework is designed around three clinically relevant objectives not simultaneously achieved by either the spine coil alone or conventional dedicated breast coils: 1) enhanced central volumetric RF sensitivity with improved signal homogeneity across the whole ROI, 2) free geometric scalability across approximately A- to DD-cup breast geometries, and 3) infrastructure-compatible operation that enables integration with standard MRI hardware without scanner-side modification and across vendor-specific receive systems. Three prototype sizes with diameters of 95, 125, and 155 mm were designed, fabricated, and validated through electromagnetic simulations and experimental MRI measurements to accommodate clinically relevant anatomical variability. Each VMR incorporates an intrinsic self-detuning mechanism and couples wirelessly to the spine coil's receive chain without requiring any scanner hardware modification.

The resulting VMR-augmented spine coil (VMR-SC) configuration transforms the spatial RF sensitivity profile of the spine coil, improving both signal penetration into the central breast-equivalent volume and spatial uniformity across the target ROI. Compared with the spine coil alone, whose high-SNR region is largely confined to the posterior side adjacent to the patient table, the VMR-SC substantially expanded high-sensitivity coverage along the vertical direction. Using SNR > 100 as the threshold for high-SNR coverage, the VMR-SC increased the vertical coverage height from approximately 27.6 mm to 139.7 mm in sagittal-plane imaging, corresponding to an average 5.09-fold expansion across the three VMR sizes. Compared with a commercial 16-channel breast coil under identical imaging conditions, the VMR-SC provided a more homogeneous signal distribution within the target imaging region, reducing the average SNR non-uniformity, quantified by the coefficient of variation (CV), from 0.195 to 0.062 across the three VMR sizes and three imaging planes; this corresponds to an approximately 68% reduction in spatial non-uniformity. Under SENSE parallel imaging at acceleration factors of 2, 3, and 4—the range routinely employed in clinical breast MRI—the VMR-SC maintained a 2- to 4-fold central ROI SNR advantage over the breast coil across all three VMR sizes, supporting its potential as a competitive wireless breast MRI receiver platform. These results demonstrate that a standard spine coil, when augmented with the VMR, can achieve breast imaging performance that comparable or even exceeds that of dedicated commercial hardware, providing a practical and accessible path toward high-quality breast MRI in resource-constrained clinical environments.

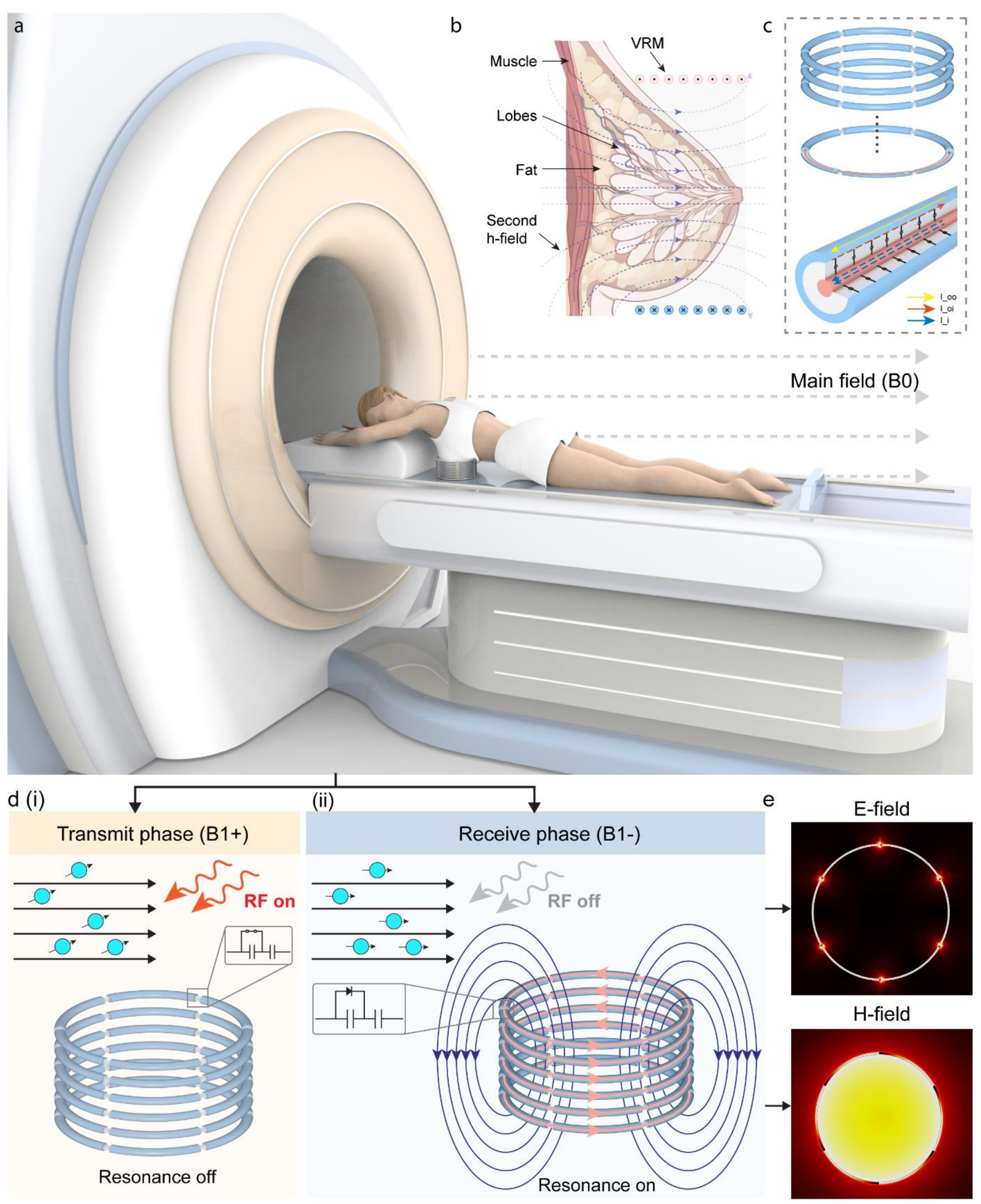

**Figure 1. Overview of the wireless volumetric metamaterial resonator–augmented spine coil (VMR-SC) platform for breast MRI. a,** Schematic illustration of breast MRI performed using the VMR-SC configuration on a conventional clinical MRI system. The patient is positioned prone, with the VMR surrounding the breast and coupling wirelessly to the built-in spine receive coil beneath the patient table. **b,** Sagittal anatomical view showing the spatial relationship among the

breast tissue and the VMR, with the direction of the main magnetic field $B_0$ indicated. **c,** Structural design of the VMR, consisting of multiple coaxially shielded resonant rings stacked along the axial direction. Each ring is assembled from segmented coaxial cable, whose inner and outer conductors provide distributed capacitance and support subwavelength resonance at the MRI Larmor frequency. **d,** Power-dependent transmit-receive operation of the VMR. During the high-power RF transmit phase (**i**), the induced voltage forward-biases the integrated PIN diodes, electrically bridging the coaxial conductors and detuning the VMR from the Larmor frequency. During the low-power receive phase (**ii**), the PIN diodes remain off, allowing the VMR to resonate and generate a secondary RF magnetic field that enhances the local $B_1^-$ sensitivity within the enclosed imaging volume. **e,** Simulated electric- and magnetic-field distributions of the receive-active VMR, showing confinement of the dominant electric field within the coaxial cable structure and concentration of the magnetic field inside the central imaging volume.

## Result

### Design strategy of the volumetric metamaterial resonator

The VMR consists of a series of coaxially shielded resonant rings stacked along the axial direction (**Fig. 1c**). Each ring functions as a unit cell and is assembled from six identical segments fabricated from commercially available coaxial cable (Alpha Wire 9849, diameter 2.54 mm). The segments are electrically connected end-to-end by soldering the inner conductor of one segment to the outer conductor of the adjacent segment, forming a closed resonant loop (Supplementary **Fig. S1**). In contrast to conventional wireless metamaterial resonators that rely on lumped capacitors [28,29,45] or high-permittivity dielectric materials [46] to achieve subwavelength resonance at the Larmor frequency, the VMR exploits the distributed capacitance inherently formed between the inner and outer conductors of the coaxial structure, together with deliberately engineered structural gaps [33,34]. This approach eliminates the lossy contributions of discrete capacitive components and promotes a higher unloaded quality factor, as confirmed by supports a sharper resonant response, as indicated by VNA measurements (Supplementary **Fig. S2**). The scalability of the VMR is governed by four geometric parameters: the inter-unit spacing (s), the number of stacked unit cells (n), the number of segments per ring (m), and the length of each segment (l). These parameters collectively determine the effective inductance and distributed capacitance of the resonator, enabling flexible tuning of both the volumetric geometry and the operating frequency. Specifically, the VMR can be geometrically scaled for different breast sizes while remaining resonant under a 3.0 T MR scanner, and can also be frequency-tuned for MRI systems operating at different field strengths, including 1.5 T, 3.0 T, and 7.0 T, as summarized in Supplementary **Table S1**.

A key requirement for the safe use of wireless resonators in MRI is transmit–receive selectivity: the resonator must be electromagnetically inactive during the high-power RF transmit phase, but active during signal reception (**Fig. 1d**). This prevents interference with the prescribed B1+ excitation pulses while preserving receive-phase signal enhancement. To enable this MRI-compatible behavior, a PIN diode (MADP-000235-10720T, MACOM Technology Solutions) is integrated into each coaxial segment to electrically bridge the inner and outer conductors. During RF transmission (**Fig. 1d (i)**), the kW-scale B1+ pulse at the Larmor frequency forward-biases the PIN diodes, effectively shorting the coaxial conductors and shifting the metamaterial resonance away from the Larmor frequency. The metamaterial is therefore detuned and remains inactive during excitation. During signal reception (**Fig. 1d (ii)**), the μW-scale MR signal is too weak to forward-bias the PIN diodes. The diodes remain effectively off, allowing the metamaterial to resonate at the Larmor frequency. In this receive-active state, induced currents flow along the coaxial segments and generate a secondary RF magnetic field that contributes to the local B1− field distribution, thereby improving receive sensitivity and enhancing the MR signal detected by the external receiver coil. As a result, integration enables a power-dependent self-detuning mechanism that automatically switches the VMR out of resonance during the high-power RF transmit phase while restoring resonance during signal reception, ensuring safe and seamless coexistence with standard MRI hardware and pulse sequences [32-34,50-54]. In addition to self-detuning, the coaxial architecture further supports safe operation by localizing the dominant resonant electric energy within the cable structure. The electric field is primarily confined to the dielectric region between the inner and outer conductors of the coaxial geometry, contributing to reduced stray electric fields in the tissue-facing region. This confinement limits capacitive coupling to adjacent tissue while preserving the magnetic response needed to support an induced B1- field within the ROI (**Fig. 1e**). As a result, the VMR minimizes undesired electric-field exposure during receive operation while maintaining efficient magnetic coupling for SNR enhancement.

To experimentally validate geometric scalability, three VMR prototypes were fabricated with outer diameters of 95 mm, 125 mm, and 155 mm, corresponding to small, medium, and large configurations for a clinical 3.0 T MR system (Philips Healthcare). These prototypes span a broad range of clinically relevant breast dimensions and confirm the robustness of the VMR design across varying geometries.

**Electromagnetic operating principle of the volumetric metamaterial resonator**

The electromagnetic behavior of the VMR originates from a current self-cancellation mechanism inherent to the coaxial geometry. At resonance, each coaxially shielded ring is excited by the incident RF electromagnetic field, inducing surface currents on both the inner and outer conductors. Owing to the skin effect, the current on the outer conductor separates into two components: a current flowing along its inner surface ($I_{oi}$) and a current flowing along its outer surface ($I_{oo}$). The current on the inner conductor ($I_i$) is confined to its outer surface, where $I_i$ flows from the open end of each segment toward the soldered junction, while $I_{oi}$ propagates in the opposite direction along the inner surface of the outer conductor (**Fig. S3**). **Fig. 2b** presents electromagnetic simulation results for the three surface current densities on the inner and outer conductors of a single unit cell, with $I_{oo}$ largely uniform throughout the unit cell, $I_i$ and $I_{oi}$ maximum at one end of the outer gap and minimum at the other end of the outer gap for each segment. The surface current strengths are then plotted in **Fig. 2c**, because $I_i$ and $I_{oi}$ flow with approximately equal amplitudes—as confirmed by the simulated surface current distributions in **Fig. 2b**—their associated magnetic fields largely cancel, leaving $I_{oo}$ as the dominant contributor to the secondary magnetic field. This self-cancellation suppresses near-field radiation from the inner current components and yields a stable, well-defined magnetic field source governed primarily by the uniformly distributed outer-surface current. Compared with unshielded loop resonators, this configuration renders the field enhancement less sensitive to geometric perturbations and tissue loading variations, contributing to the robustness of the VMR response.

The axial distribution of the dominant outer-conductor current $I_{oo}$, was further quantified across a 21-ring VMR (**Fig. 2d**). For each rung, $I_{oo}$ was extracted from the corresponding resonant ring after excluding the gap region, with the marker representing the mean current amplitude and the error bar indicating its local fluctuation along the ring. This approximately sinusoidal current profile, with central rings carrying the highest current magnitude and amplitude tapering toward the outer rings, suggests that the stacked rings operate as a collectively coupled resonant system rather than as independent resonators. The weaker response at the two ends is consistent with edge effects and reduced axial coupling, whereas the central rings experience stronger bidirectional coupling and therefore support the largest resonant currents. Such an ordered current envelope indicates a coherent volumetric resonance (so-called co-rotating mode), which is essential for producing a stable and spatially extended magnetic-field response within the VMR. Operating at the co-rotating mode produces a magnetic field distribution that is strongly concentrated at the center of the volumetric ROI.

Despite this collectively coupled mode, inductive coupling between adjacent rings gives rise to multiple collective eigenmodes, each characterized by a distinct current distribution and magnetic field pattern, identifiable as separate resonance dips in the reflection spectrum. **Fig. 2e** presents the experimentally measured $S_{11}$ of a 21-ring VMR, revealing three distinct resonant modes measured using a Vector Network Analyzer (VNA, P5020B, Keysight Inc.; setup shown in Supplementary **Fig. S4**). The simulated current distributions for each mode are shown in the inset of **Fig. 2e**. Mode 1, referred to as the co-rotating mode, exhibits an in-phase current distribution across all rings, forming a fully constructive collective excitation. Modes 2 and 3 display partially non-constructive current patterns that result in less efficient and less uniform magnetic field concentration. The measured modal behavior is consistent with eigenvalue calculations based on coupled resonator dynamics (**Supplementary Note S1**).

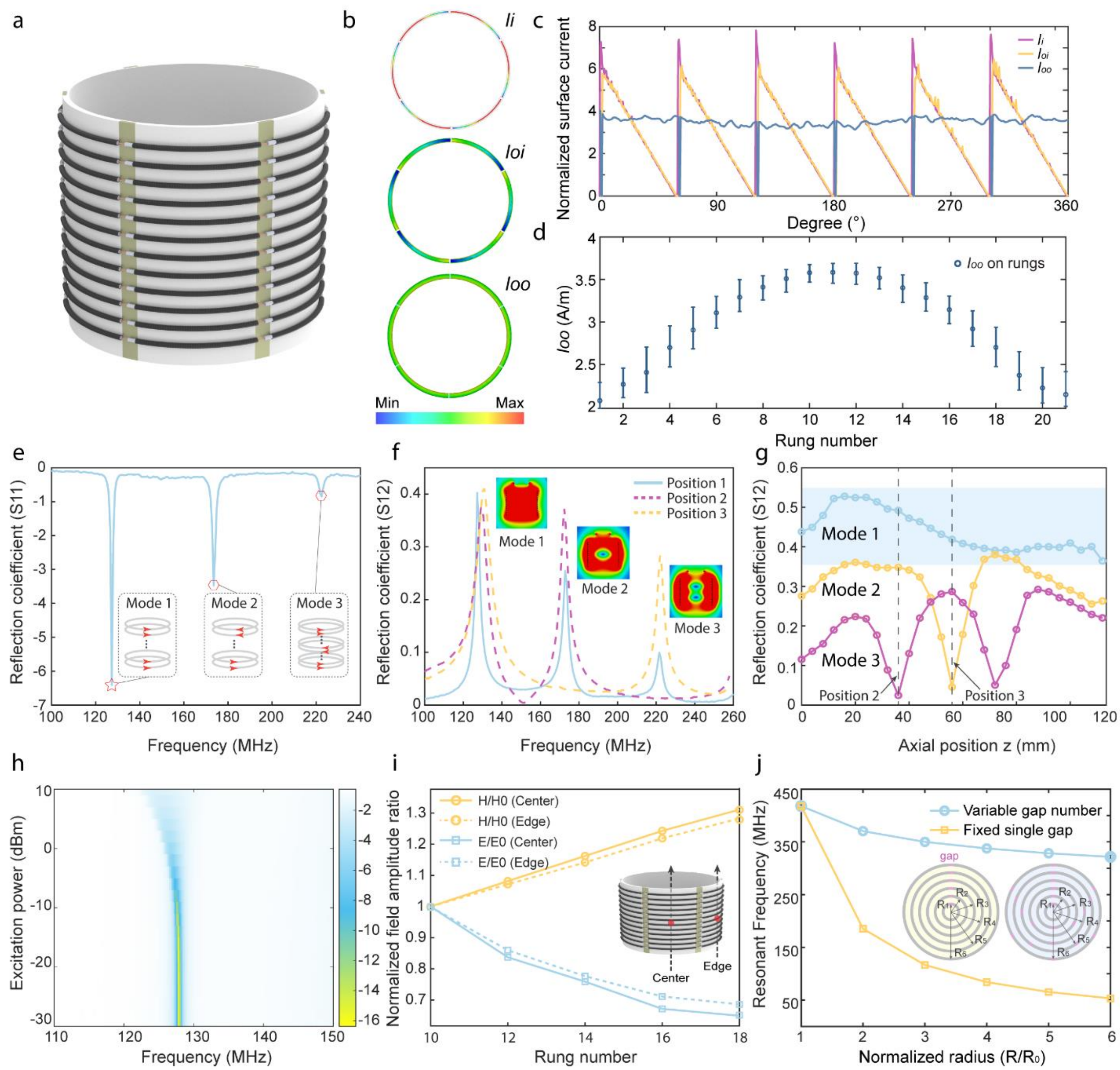

**Figure 2. Electromagnetic mechanism and geometric scalability of the VMR. a,** Three-dimensional model of the axially stacked VMR, comprising coaxially shielded resonant rings supported by a 3D-printed holder. **b,** Simulated surface-current distributions on a representative resonant ring, including the current on the outer surface of the inner conductor ($I_i$), the current on the inner surface of the outer conductor ($I_{oi}$), and the current on the outer surface of the outer conductor ($I_{oo}$). **c,** Normalized amplitudes of $I_i$, $I_{oi}$, and $I_{oo}$ along the circumference of the segmented ring. The approximately equal and oppositely directed $I_i$ and $I_{oi}$ components produce substantial magnetic-field cancellation, leaving $I_{oo}$ as the dominant contributor to the external RF magnetic field. **d,** Axial distribution of $I_{oo}$ across the resonant rings of a 21-ring VMR. Markers indicate the mean current amplitude on each ring after excluding the gap regions, and error bars

indicate the local current variation along the ring. The centrally peaked, gradually tapered profile reflects coherent collective coupling among the stacked rings. **e,** Experimentally measured reflection spectrum ($S_{11}$) of the 21-ring VMR, showing three collective resonant modes. Insets show the corresponding simulated ring-current patterns. Mode 1 exhibits an in-phase co-rotating current distribution, whereas modes 2 and 3 contain partially non-constructive current patterns. **f,** Transmission spectra ($S_{12}$) measured with one loop antenna fixed below the VMR and a second antenna positioned at three representative locations along the central axis. Insets show the simulated magnetic-field distributions of the three modes. **g,** Transmission strength associated with modes 1-3 as a function of axial antenna position. Mode 1 remains continuously detectable across the scanned range, whereas modes 2 and 3 exhibit axial transmission nulls associated with higher-order modal nodes and destructive interference. **h,** Power-dependent reflection spectrum showing progressive displacement of the VMR resonance as the VNA excitation power increases, confirming the nonlinear self-detuning response enabled by the integrated PIN diodes. **i,** Simulated H- and E-field amplitudes at the VMR center and tissue-facing edge as the number of stacked rings increases. Values are normalized to those of the 10-ring VMR at the corresponding sampling locations. Increasing the number of rings strengthens the magnetic response while reducing the electric-field amplitude. **j,** Simulated resonance frequency as a function of normalized ring radius for fixed single-gap and variable-gap designs. Increasing the number of structural gaps maintains a more stable resonance frequency as the ring diameter increases, enabling geometric scaling of the VMR while preserving operation near the target Larmor frequency.

To experimentally probe the axial modal response, $S_{12}$ was measured using a vector network analyzer, with one antenna fixed above the VMR and the other translated along the central axis of the structure. The transmission coefficient spectra measured at three representative axial positions are shown in **Fig. 2f**. While the resonances associated with modes 2 and 3 undergo pronounced amplitude variations and spectral reorganization as the receive antenna is translated, the resonance corresponding to mode 1 remains clearly identifiable and maintains a strong transmission response. These observations suggest that mode 1 is significantly more robust to axial displacement than the higher-order modes. Insets in Fig. 2f show the simulated magnetic field distributions associated with the three eigenmodes.

To quantify the axial evolution of each eigenmode, the transmission minima associated with the three resonances were extracted from all measured positions and plotted as a function of axial position in **Fig. 2g**. Mode 1 maintains a continuous response across the entire 112-mm scanning range and exhibits only moderate variation in transmission strength. In contrast, modes 2 and 3 display pronounced transmission nulls at specific axial locations, indicating the presence of axial nodes and destructive modal interference. These null positions coincide with the phase discontinuities expected for higher-order collective excitations. The absence of such nodes in mode 1 confirms that all resonant rings oscillate in phase, corresponding to a fully constructive collective excitation state with coherent magnetic-field reinforcement throughout the structure.

The co-rotating mode 1 produces a strongly concentrated and volumetrically uniform magnetic field, whereas modes 2 and 3 generate less coherent field patterns with reduced central enhancement. The VMR is therefore designed and operated at mode 1 by tuning the resonant frequency for mode 1 to align with the Larmor frequency. This current profile supports a centrally peaked magnetic field distribution that remains consistent across varying numbers of stacked rings, as demonstrated by simulated field distributions under co-rotating mode excitation for arrays of different ring counts (Supplementary **Fig. S5**). This robustness confirms that the centrally concentrated sensitivity profile is a fundamental characteristic of the co-rotating mode rather than a geometry-specific artifact, enabling RF field penetration into deep tissue regions that planar resonator geometries cannot reach. Unlike planar resonators whose sensitivity decays rapidly away from the structure, the VMR achieves volumetric field enhancement across the entire ROI with a pronounced concentration in the central, deepest region—a characteristic directly relevant to the diagnostic requirements of breast MRI. A quantitative comparison of axial field profiles between a planar ring array and the VMR is provided in Supplementary **Fig. S6**, demonstrating the dark-spot artifact and limited penetration depth inherent to planar configurations.

**Compatibility and safety characterization**

The VMR couples to the spine coil wirelessly through inductive near-field interaction, requiring no electrical connection or hardware modification to the scanner. When positioned coaxially around the breast, the VMR resonates at the Larmor frequency and generates a secondary RF magnetic field that constructively enhances the $B1^-$ sensitivity within the enclosed volume. This enhanced sensitivity is captured by the underlying spine coil receive channels, which remain the active signal pathway to the scanner. The relative axial alignment and coaxial overlap between the VMR and the spine coil determine the strength of inductive coupling and were characterized through simulation and bench measurements prior to MRI experiments (Supplementary **Fig. S7**). Under optimized positioning, the VMR-SC configuration reshapes the spatial $B1^-$ sensitivity profile of the spine coil, extending effective sensitivity into the central and anterior breast regions that are poorly covered by the spine coil alone.

The VMR incorporates a self-activated detuning mechanism that separates its behavior during the transmit and receive phases of MRI acquisition. During the transmit phase, the high-power $B1^+$ field forward-biases the PIN diodes integrated into each coaxial segment, shifting the resonant frequency away from the Larmor frequency and suppressing resonant enhancement. During the receive phase, the substantially lower $B1^-$ field is insufficient to forward-bias the diodes, and the VMR returns to resonance. This power-dependent switching is entirely self-activated, requiring no external control circuitry or synchronization with the scanner. **Fig. 2h and Fig. S8** show the measured S11 frequency as a function of VNA excitation power, confirming progressive resonance displacement as power increases to levels representative of clinical transmit fields.

Two complementary experiments were performed to verify that the VMR does not perturb the $B1^+$ distribution during the transmit phase. First, the mean SNR within a homogeneous phantom ROI was measured across flip angles (FA) ranging from 10° to 90° for the VMR–SC, spine coil alone, and birdcage coil alone configurations (**Fig. S9**) [55]. All three configurations follow an identical flip-angle-dependent trend, confirming that the VMR does not measurably alter global transmit behavior. Second, B1+ maps were acquired using a dual-TR method to provide a spatially resolved, quantitative assessment of transmit field uniformity in the presence of the VMR (**Fig. S10**). No spatial distortion of the $B1^+$ distribution attributable to the VMR was observed. Together, these experiments establish transmit field compatibility at both the global and spatially resolved levels, confirming that the VMR requires no recalibration of RF power settings, pulse sequence parameters, or SAR constraints when integrated into standard clinical imaging protocols.

RF safety was evaluated through electromagnetic simulation and experimental MR thermometry. Local specific absorption rate averaged over 10 g of tissue ($SAR_{10g}$) was simulated using a modified full-body human model under 1 W normalized forward input power, with and without the VMR present. Experimental validation was performed using MR thermometry based on the proton resonance frequency shift method during high-power scanning sequences, monitoring temperature changes for the spine coil alone and VMR-SC configurations (Supplementary **Fig. S11**). All configurations exhibited comparable phase evolution with no evidence of localized heating attributable to the VMR, confirming consistency with the simulated SAR results and establishing the safety of the VMR under clinical operating conditions.

The selective transmit–receive response is necessary for compatibility with modern MRI scanners and for safe operation, but it is not sufficient on its own. RF-induced heating is primarily associated with electric-field-driven conductive losses in tissue. Even during the low-power receive phase, the magnetic and electric field components inevitably coexist as part of the resonant electromagnetic response. Although the electric field cannot be completely eliminated, a carefully engineered resonator architecture can substantially suppress stray electric fields at the tissue-facing surface. To further evaluate the field-confinement advantage of the coaxial VMR architecture, we simulated VMRs with different numbers of stacked resonant rings and compared their H- and E-field responses after normalization to the initial 10-ring design, where $H_0$ and $E_0$ denote the field amplitudes of the 10-ring VMR at the same center or edge location (**Fig. 2i**). Two representative locations were selected for analysis: the geometric center of the VMR and an edge position near the tissue-facing sidewall. As the ring number increased, the normalized H-field amplitude monotonically increased at both the center and edge locations, whereas the normalized E-field amplitude progressively decreased. This opposite trend indicates that increasing the axial stacking density strengthens the receive-relevant magnetic response while suppressing stray electric fields near the tissue-facing region, supporting the VMR's ability to enhance magnetic coupling while reducing undesired capacitive interaction with adjacent tissue. This field-confinement mechanism was further confirmed by an axial line-profile analysis passing through the coaxial cable cores, which showed strong E-field localization inside the cable structure and negligible field leakage outside the cables (Supplementary **Fig. S12**).

The scalability of the VMR is enabled by gap engineering on each resonator ring, which compensates for changes in resonance induced by increasing ring diameter. A quantitative analysis of the relationship between resonance frequency and ring diameter with and without gap-number tuning is shown in **Fig. 2j**. Six linearly increased diameters were investigated. Compared with a fixed single-gap configuration, increasing the number of gaps divides the enlarged ring into additional cable segments, thereby maintaining a comparable segment length and effective electromagnetic response as the ring diameter increases. With engineered electrical bridging across these gaps, the variable-gap configuration preserves relatively stable resonance frequencies across a wide range of ring diameters. The remaining minor frequency shifts can be further compensated by adjusting the inter-unit spacing (s) or the number of stacked unit cells (n). Based on this tuning strategy, three VMR prototypes were fabricated to represent small, medium, and large effective imaging volumes for MRI validation.

## On-bench Validation

### Phantom validation

Phantom experiments were designed to systematically evaluate three aspects of VMR–SC performance: SNR enhancement relative to both the spine coil and the 16-channel breast coil, geometry-enabled SNR compensation, and compatibility with parallel imaging under clinically relevant acceleration factors.

Three agarose gel phantoms (1 wt%) with diameters matched to the three VMR sizes (95, 125, and 155 mm) were imaged on a clinical 3.0-T MRI system. Three configurations were compared: the VMR-SC, the spine coil alone, and a commercial 16-channel breast coil. These comparisons allow the metamaterial-induced signal enhancement to be isolated relative to the baseline spine coil, while also benchmarking the performance of the VMR-SC configuration against a dedicated clinical receiver array. For the VMR-SC and spine-coil-only measurements, each phantom was positioned at the same location relative to the spine coil to ensure identical loading and imaging geometry. For the breast coil measurements, each phantom was placed within one chamber of the breast coil housing, and the movable baffle was adjusted to the closest geometric fit for each phantom diameter. All three configurations were imaged in axial, sagittal, and coronal planes for each phantom size, enabling comparison of signal distribution across all three spatial dimensions.

Representative SNR maps and qualitative signal distributions are shown in **Figs. 3a-c** for each size. The VMR-SC configuration produces a brighter and more spatially extended signal distribution throughout the effective coverage region of the VMR, as indicated by the dashed rectangle. Within this region, the resonant structure concentrates and redistributes the RF magnetic field toward the interior of the imaging volume, thereby increasing receive sensitivity in regions that are typically weakly detected by surface coils. The spine-coil-only configuration exhibits limited sensitivity, primarily confined to regions immediately adjacent to the coil surface where the phantom contacts the patient table. The breast coil configuration shows signal enhancement localized near the embedded coil elements in the housing walls, central baffle, and bottom elements—a distribution that reflects the fixed geometry of the receive array rather than the anatomy of the target. By adjusting the relative axial position between the VMR and the phantom, this enhanced region can also be translated along the imaging volume, providing additional flexibility in selecting the target coverage region.

Quantitative SNR profiles extracted along the central axes confirm the volumetric advantage of the VMR-SC. **Figure 3d** shows SNR profiles along the x-direction from the axial plane, and **Figure 3e** shows profiles along the y-direction from the sagittal plane, for all three VMR sizes. Along both directions, the VMR-SC provides substantial signal enhancement relative to the spine coil alone across a large portion of the imaging volume. When compared with the breast coil, the VMR-SC produces comparable or higher SNR throughout much of the interior region, with the exception of peripheral zones immediately adjacent to the breast coil housing elements. To visualize the spatial distribution of this difference, two-dimensional ΔSNR maps ($\Delta SNR = SNR_{VMR\text{-}SC} - SNR_{BR\ only}$) were computed between the VMR-SC and the breast coil (**Fig. 3g**). Positive ΔSNR values occupy, on average, 69.3 % of the imaging region in the axial planes and 91.43 % in the sagittal planes across the three VMR sizes on the whole phantom; it is also noticed that, the VMR does not designed to cover the whole phantom, in the enclosed volumetric region by VMR, the positive ΔSNR portion is even more pronounced.t In both cases, the enhancement is most pronounced in the central region of the phantom, consistent with the centrally peaked magnetic-field distribution of the co-rotating mode established in the electromagnetic characterization above. **Figure 3f** shows the SNR profiles along the z-direction extracted from the coronal plane. The three configurations exhibit similar spatial profiles, while the VMR-SC configuration consistently provides the highest SNR across all positions.

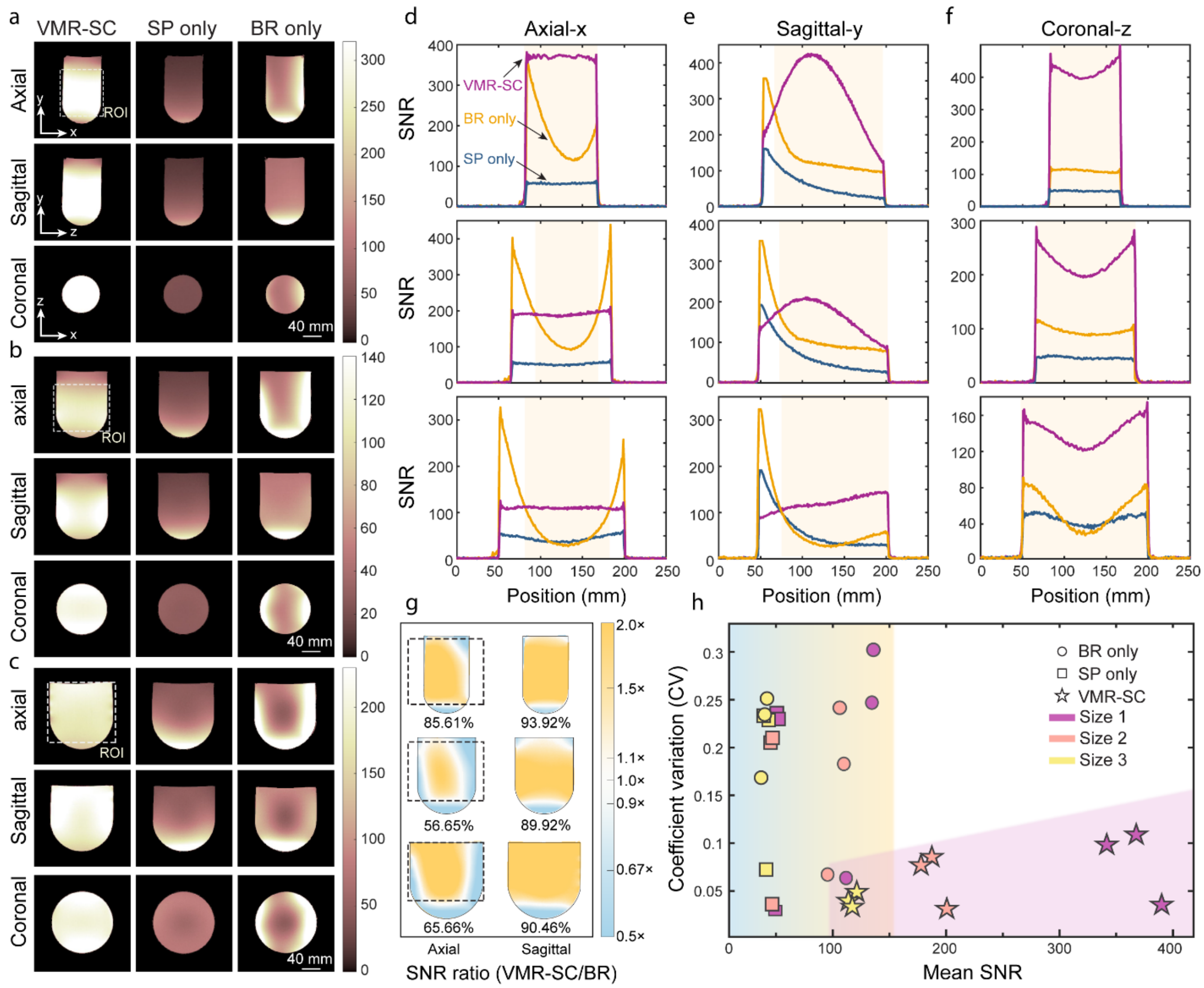

**Figure 3. Phantom validation of SNR enhancement and spatial uniformity provided by the VMR-SC across three imaging volumes. a-c,** SNR maps of the small, medium, and large phantoms, respectively, acquired using the VMR-SC, spine coil alone (SP only), and commercial 16-channel breast coil alone (BR only) in the axial, sagittal, and coronal imaging planes. Dashed rectangles indicate, for spatial reference, the projected extent of the volume enclosed by the VMR in each imaging plane. Coordinate directions are indicated in **a**, and scale bars represent 40 mm. **d-f,** SNR profiles extracted along the x-direction in the axial plane (**d**), y-direction in the sagittal plane (**e**), and z-direction in the coronal plane (**f**). The top, middle, and bottom rows correspond to the small, medium, and large phantoms, respectively. Shaded regions indicate positions where the VMR-SC provides higher SNR than the breast coil. **g,** Spatial maps of the SNR ratio between the VMR-SC and breast coil configurations ($SNR_{VMR-SC}/SNR_{BR}$) in the axial and sagittal planes for the three phantom sizes. Ratio values greater than 1 indicate regions in which the VMR-SC provides higher SNR than the breast coil. Dashed rectangles indicate the VMR-covered regions for spatial reference, whereas the percentages denote the fractions of pixels across the entire sagittal phantom cross-section for which $SNR_{VMR-SC}/SNR_{BR} > 1$. **h,** Mean SNR plotted against

the coefficient of variation (CV) for all three receiver configurations, phantom sizes, and imaging planes. Marker shapes denote the receiver configuration, and colors denote the phantom size. The VMR-SC measurements predominantly occupy the high-mean-SNR, low-CV region, demonstrating improved signal strength and spatial uniformity compared with the spine coil and commercial breast coil alone.

To summarize overall imaging performance across all sizes and imaging planes, **Figure 3h** presents a scatter plot of mean SNR versus coefficient of variation (CV) for the three configurations, with each marker representing one combination of device size (small, medium, and large) and imaging plane (axial, sagittal, and coronal), yielding nine measurements per configuration. The VMR-SC consistently occupies the lower-right region of the plot, corresponding to higher mean SNR combined with lower spatial variation—indicating a more favorable balance between signal strength and uniformity than either alternative. The breast coil measurements exhibit larger CV values despite moderate SNR levels, reflecting the spatially non-uniform sensitivity distribution inherent to its fixed-geometry receive array. The spine coil alone produces substantially lower SNR across all measurements.

It is worth noting that the decrease in mean SNR with increasing VMR diameter, visible in **Fig. 3h**, reflects a controlled prototype design rather than a fundamental limitation of the VMR framework. In the three devices studied here, the number of segments per resonator ring was intentionally kept constant (m=6) across all diameters to isolate the diameter-dependent scaling effect from other geometric variables. As the resonator diameter increases, the distributed capacitance of the coaxial segments increases, lowering the intrinsic resonant frequency of each ring. To maintain resonance at the Larmor frequency, the effective inductance of the resonator array must therefore be reduced. In the present prototypes, this was achieved by increasing the inter-unit spacing, (s), which consequently reduces the number of stacked resonator units, (n), within the same axial height. Under this constraint, (n) decreases by nearly a factor of two at each scaling step as the device diameter increases from 95 mm to 125 mm and to 155 mm, leading to the corresponding decrease in achievable SNR observed in Fig. 3a-c. Importantly, this trend does not represent an intrinsic performance ceiling of the VMR architecture. Rather, it reflects the specific choice to keep (m) fixed during diameter scaling. As demonstrated in Supplementary **Fig. S13**, reducing the segment length at a fixed diameter of 125 mm increases the effective resonator density and partially restores SNR enhancement without compromising volumetric field coverage. These results indicate that the VMR should not be regarded as a single fixed geometry, but as a scalable resonator platform whose segment length, ring segmentation, and inter-unit spacing can be systematically co-optimized across different anatomical dimensions. Building on this geometric flexibility, the VMR framework enables geometry-enabled SNR compensation, allowing the electromagnetic performance to be maintained across different anatomical sizes. This capability contrasts markedly with the size-dependent performance variation observed in conventional coil designs. For example, although the commercial breast coil incorporates a movable baffle to accommodate different breast sizes, the positions of its receive elements remain fixed. Consequently, the electromagnetic coupling between the coil and the imaging target varies with phantom diameter, leading to systematic SNR differences across sizes. In contrast, the VMR allows the resonator geometry to be directly tuned to compensate for these scaling effects. By reducing the segment length and thereby increasing the effective resonator density, the SNR reduction associated with larger device diameters can be partially recovered. After this geometry-enabled SNR compensation, the VMR-SC configuration exhibits markedly improved electromagnetic performance consistency across phantoms representing different target body sizes. **Fig. S14** summarizes this effect, showing that both the SNR profiles and the CV are improved after compensation. These results highlight the structural advantage of geometry-specific resonator tuning in the VMR framework compared with the purely mechanical size adaptation used in conventional coil designs.

**Parallel imaging performance validation**

Having established the SNR and uniformity advantages of the VMR-SC under unaccelerated conditions, we next evaluated whether the metamaterial structure preserves parallel imaging capability—a critical requirement for clinical deployment. Parallel imaging accelerates MRI acquisition by exploiting spatial variations in receiver coil sensitivity profiles to reconstruct undersampled k-space data [56]. A passive wireless resonator that alters the spatial sensitivity encoding of the underlying receive array could degrade parallel imaging reconstruction quality, even while improving raw SNR. It is therefore necessary to verify that the VMR enhances signal magnitude without disrupting the spatial encoding characteristics of the spine coil array.

Parallel imaging experiments were performed on the same 3.0-T MRI system using SENSE reconstruction at acceleration factors of 2, 3, and 4—the range routinely employed in clinical breast MRI. Representative images acquired with the spine coil alone, breast coil alone, and the VMR-SC at each acceleration factor are shown in **Figs. 4a-c**, with the dashed rectangle indicating the VMR-enclosed region. The introduction of the VMR produces a substantially brighter signal throughout the ROI at all acceleration factors. Quantitative SNR enhancement ratios calculated along the horizontal and vertical center lines of the ROI further clarify how the signal distribution is modified (**Figs. 4d-f and Fig. S15**). Along the horizontal direction, compared to the spine coil only configuration, the VMR-SC's enhancement ratio remains broadly elevated, reaching approximately 8.3-fold for the small VMR, 4.6-fold for the medium, and 3.3-fold for the large. Along the vertical direction, the VMR-SC redistributes the receive sensitivity deeper into the phantom, with the enhancement reaching approximately 8.2-fold at AF = 2 and 7.8-fold at AF = 4 in the central region of the small VMR, and remaining substantial in the medium and large VMRs with peak values of approximately 4.3-fold and 4.5-fold, respectively. The average VMR-SC/SP enhancement across the whole phantom was 6.7-fold, 3.7-fold, and 2.9-fold for the small, medium, and large VMRs, respectively, confirming that the VMR amplifies the weak receive sensitivity of the spine coil under accelerated imaging conditions.

Benchmarked against the 16-channel breast coil, the VMR-SC provides higher and more uniform SNR within the VMR-covered central imaging region while avoiding the strongly peripheral signal distribution imposed by the fixed breast-coil geometry. The VMR-SC/BR enhancement reaches approximately 3.8-4.0-fold along the horizontal profiles and approximately 3.0-4.0-fold along the vertical profiles, depending on VMR size, position, and acceleration factor. This advantage is maintained across AF = 2-4, indicating that the VMR-SC preserves robust SNR enhancement under clinically relevant parallel imaging conditions. The breast coil retains a local SNR advantage only in peripheral regions immediately adjacent to its housing elements, a pattern consistent with the unaccelerated results.

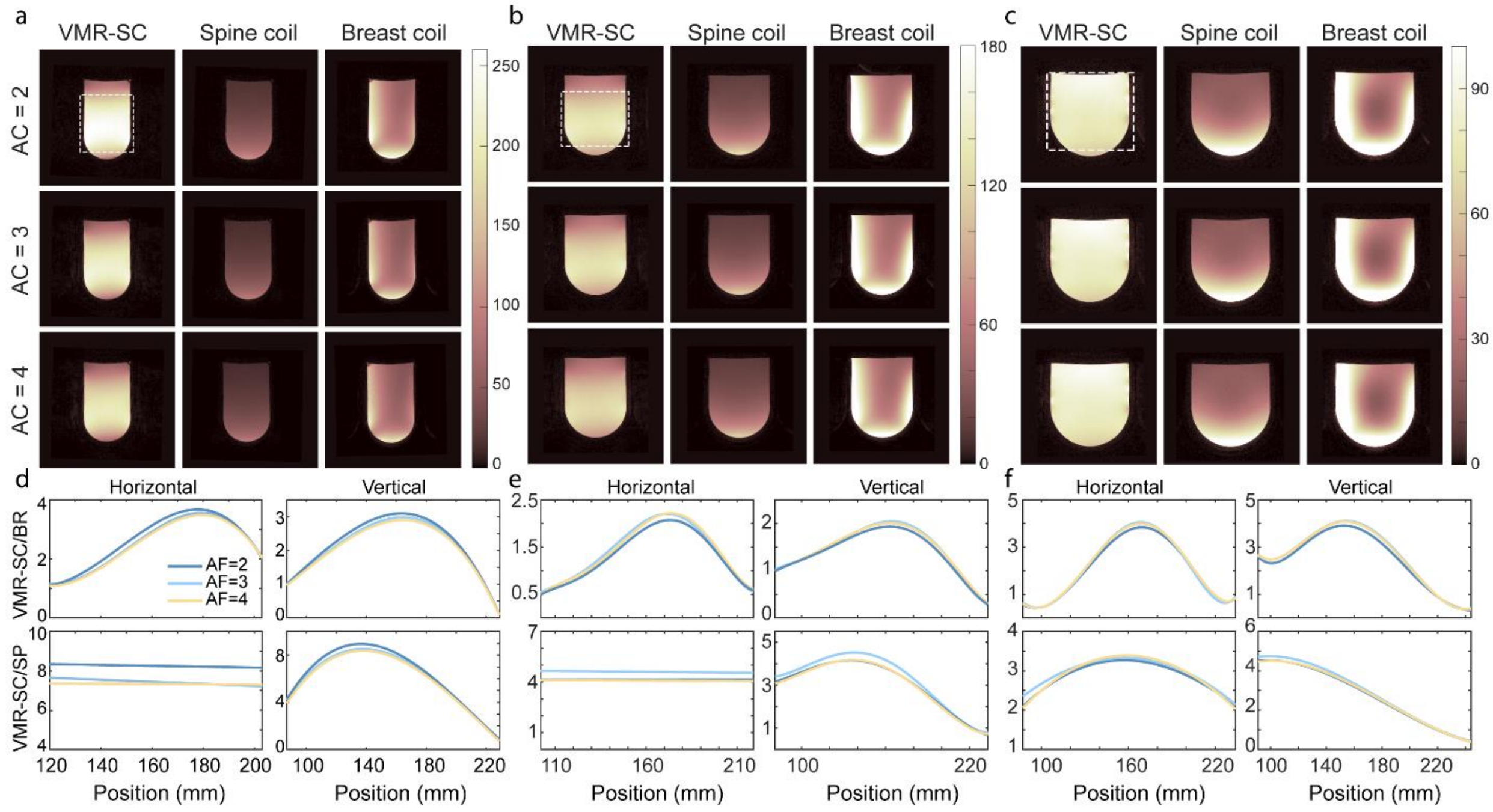

**Figure 4. SENSE parallel imaging performance of the VMR-SC across different device sizes and acceleration factors. a-c,** Sagittal SNR maps of the small (**a**), medium (**b**), and large (**c**) phantoms acquired using the VMR-SC, spine coil alone, and commercial 16-channel breast coil alone at acceleration factors (AFs) of 2, 3, and 4. Dashed rectangles mark, for spatial reference, the projected extent of the volume enclosed by the corresponding VMR. **d-f,** SNR enhancement ratios for the small (**d**), medium (**e**), and large (**f**) phantoms, extracted along the central horizontal and vertical profiles. The upper row shows the enhancement relative to the breast coil, $SNR_{VMR-SC}/SNR_{BR}$, and the lower row shows the enhancement relative to the spine coil, $SNR_{VMR-SC}/SNR_{SP}$. Curves correspond to AFs of 2, 3, and 4. Ratios greater than 1 indicate higher SNR with the VMR-SC than with the corresponding reference coil.

## Ex vivo validation in a heterogeneous biological sample

To further examine whether the VMR-SC enhancement extends beyond homogeneous phantoms, we performed ex vivo MRI of a heterogeneous biological sample using clinically relevant T1wTSE and T2wTSE sequences (**Fig. 5**). The sample was imaged using the 16-channel breast coil alone and the VMR-SC configuration under otherwise identical acquisition conditions (Supplementary **Table S2**). Compared with the breast coil, the VMR-SC produced visibly higher signal intensity in both T1wTSE and T2wTSE images, while preserving the internal structural features of the sample. In particular, the VMR-SC enhanced signal visibility across the central region, where receive sensitivity is typically weaker for fixed surface-array geometries. The improved depiction of internal contrast patterns in this structured sample supports the ability of the VMR to redistribute receive sensitivity toward the interior of a non-uniform imaging volume. The ex vivo experiment provides an additional demonstration that the VMR-enhanced receive field can improve image visibility in a heterogeneous biological target beyond uniform phantom conditions.

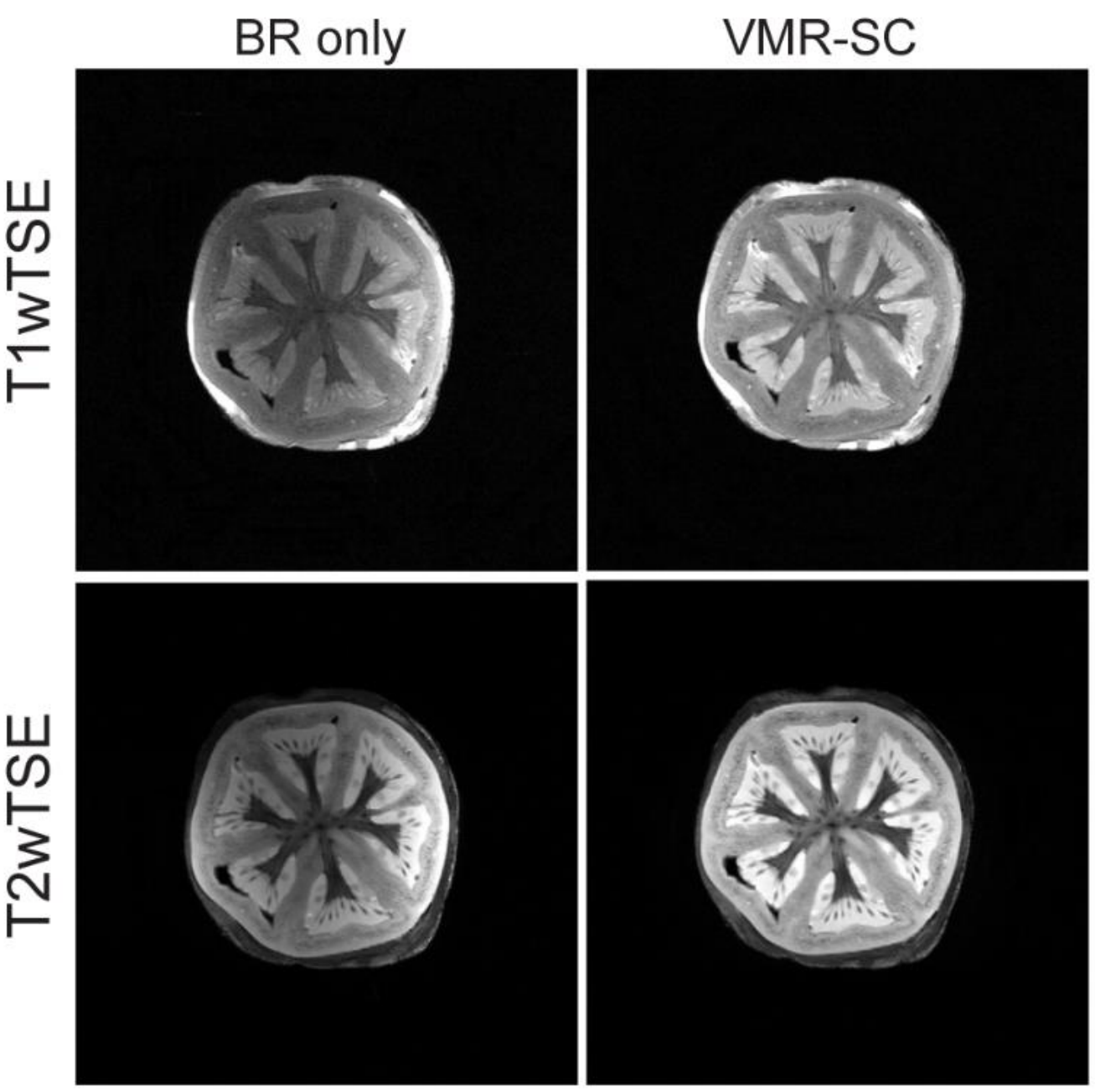

**Figure 5. Ex vivo validation of the VMR-SC in a heterogeneous biological sample.** Representative T1-weighted turbo spin echo (T1wTSE) and T2-weighted turbo spin echo (T2wTSE) images of a tomato sample acquired using the BR only and the VMR-SC under otherwise identical imaging conditions. The VMR-SC provides higher signal intensity, particularly within the central region, while preserving the internal structural features of the sample. Detailed acquisition parameters are provided in Supplementary Table S2.

## Discussion

This work presents a wireless volumetric metamaterial resonator that repurposes the built-in, multi-channel spine coil—an otherwise underused but high-performance receive array available in nearly all clinical MRI systems—into a breast imaging platform, addressing a key hardware mismatch between widely available scanner infrastructure and the anatomical requirements of high-quality breast imaging. Although dedicated breast coils offer a degree of mechanical adjustability, they remain costly, vendor-specific, and constrained by fixed receive-array geometries. In contrast, the VMR reshapes the local RF receive sensitivity through a passive, scalable volumetric resonant structure placed around the imaging target. This design enables the spine coil to provide enhanced sensitivity across the enclosed breast-equivalent volume without direct electrical connection, scanner modification, or sequence-specific hardware control. In phantom experiments, the VMR-SC configuration produced higher central SNR than a commercial 16-channel breast coil and substantially improved spatial uniformity within the target region, demonstrating that a wireless resonant insert can provide not only local signal gain but also volumetric redistribution of receive sensitivity.

A central feature of the VMR is its coaxially shielded, axially stacked architecture. The distributed capacitance of the coaxial cable supports resonance at the Larmor frequency without relying on lossy lumped capacitors, while the stacked-ring configuration forms a collective co-rotating mode that concentrates the RF magnetic response toward the interior of the imaging volume. This distinguishes the VMR from planar metamaterial or surface-resonator designs, whose sensitivity enhancement typically decays rapidly away from the device surface. In parallel, the coaxial geometry confines the dominant electric field within the cable dielectric, reducing tissue-facing stray electric fields while preserving the outer-conductor current as the dominant source of the receive-relevant magnetic response. Together, these electromagnetic features allow the VMR to increase magnetic coupling within deep regions of the target without introducing substantial capacitive interaction at the tissue-facing surface. The VMR framework also provides geometric scalability, an important requirement for breast MRI because target anatomy varies substantially across individuals. In the present prototypes, the number of segments per resonator ring was intentionally fixed to isolate the effect of diameter scaling, which led to a reduction in resonator density and a corresponding decrease in SNR for larger devices. However, this trend reflects the controlled design of the prototype set rather than an intrinsic limitation of the platform. The VMR provides multiple geometric parameters—including segment length, ring segmentation and inter-unit spacing—that can be co-optimized to preserve resonance and recover SNR across different anatomical sizes. This capability differs fundamentally from conventional coil-size adaptation, in which the housing may be mechanically adjusted while the receive-element positions remain fixed.

By contrast, the VMR enables electromagnetic tuning of the resonant structure itself, allowing the effective sensitivity distribution to be matched to the imaging volume.

MRI compatibility is essential for any wireless resonator intended for clinical translation. The VMR addresses this requirement through a PIN-diode-based self-detuning mechanism that suppresses resonance during the high-power transmit phase and restores resonance during signal reception. Bench measurements, flip-angle-dependent imaging, B1+ mapping, SAR simulations and MR thermometry collectively support compatibility with standard 3.0 T imaging conditions. These features allow the VMR to operate as a passive, self-regulating receive-enhancement structure rather than an actively controlled coil element. The VMR-SC configuration further maintained its performance under SENSE parallel imaging at acceleration factors of 2-4. This result is important because accelerated breast MRI is clinically valuable for reducing scan time, but it requires stable and spatially informative receive sensitivity profiles. A passive resonant structure that substantially distorts the encoding field could degrade reconstruction quality even if it improves raw signal intensity. The observed SNR enhancement across acceleration factors indicates that the VMR increases receive sensitivity while preserving sufficient spatial encoding compatibility with the underlying spine coil array. This finding supports the potential use of the VMR-SC configuration in clinically relevant accelerated imaging workflows. This study has several limitations. The present prototypes were optimized for proof-of-concept validation at 3.0 T, and further engineering will be needed to refine patient comfort, positioning reproducibility, manufacturability and integration with clinical breast MRI workflows. Moreover, although the VMR showed compatibility with SENSE acceleration, future work should quantify reconstruction performance more comprehensively using g-factor analysis, artifact metrics and multi-slice or volumetric acquisitions. Finally, the scalability strategy should be further expanded to include patient-specific or size-adaptive designs that preserve both resonance and resonator density across a broader range of anatomical dimensions.

Overall, this work establishes the VMR as a scalable wireless hardware platform for enhancing breast MRI using existing clinical scanner infrastructure. By combining volumetric RF field redistribution, self-detuned MRI compatibility, electric-field confinement, and parallel imaging preservation, the VMR provides a practical route toward high-SNR, more homogeneous breast imaging without relying exclusively on dedicated receive coils. This approach could support broader access to high-quality breast MRI in settings where dedicated breast coil hardware is unavailable, limited, or difficult to adapt to individual anatomy.

## Methods

### Volumetric metamaterial resonator construction

The volumetric metamaterial resonators (VMRs) were constructed from non-magnetic coaxial cable segments (9849, Alpha Wire) arranged into axially stacked resonant rings. The coaxial cable consisted of an inner conductor, dielectric insulation, outer conductor and external jacket, with diameters of 0.48 mm, 1.42 mm, 1.93 mm and 2.54 mm, respectively. Each resonant ring was assembled from ($m$) coaxial cable segments connected end-to-end by soldering the inner conductor of one segment to the outer conductor of the adjacent segment, forming a closed resonant loop. Structural gaps were introduced by selectively removing the outer conductor over a designed 4-mm gap length, while maintaining the distributed capacitance between the inner and outer conductors along the cable segments. PIN diodes (MADP-000235-10720T, MACOM Technology Solutions) were integrated between the inner and outer conductors to enable power-dependent self-detuning during MRI operation.

Three VMR prototypes were fabricated to represent small, medium, and large breast-equivalent imaging volumes. The outer diameters of the three devices were 95 mm, 125 mm, and 155 mm, respectively. For the controlled diameter-scaling study, the number of segments per resonant ring was fixed at ($m = 6$), while the inter-unit spacing, (s), and the number of stacked unit cells, (n), were adjusted to maintain resonance at the 3.0 T Larmor frequency. The final geometric parameters of each prototype, including (m), (n), (s), segment length (l), outer diameter, and effective coverage height, are summarized in **Supplementary Table S3**.

### Electromagnetic simulations

Electromagnetic simulations were performed using CST Microwave Studio Suite 2021. Magnetic-field, electric-field, and surface-current distributions were calculated using the frequency-domain solver. The simulated geometries matched the experimentally fabricated VMR prototypes unless otherwise specified. PIN diodes were modeled as 1.2-pF capacitors in the receive state and as short-circuited paths in the transmit state, consistent with established approximations for passive detuning analysis in metamaterial MRI studies.

To analyze the electromagnetic operating mechanism, surface currents on the inner conductor, the inner surface of the outer conductor, and the outer surface of the outer conductor were extracted separately and denoted as Ii, Ioi, and Ioo, respectively. The axial distribution of the dominant outer-conductor current Ioo was evaluated in a 21-ring VMR. For each rung, Ioo was extracted from the corresponding resonant ring after excluding the gap region. The marker represents the mean current amplitude, and the error bar represents the local fluctuation of Ioo along the ring.

To evaluate field confinement, VMRs with different numbers of stacked rings were simulated, and the H- and E-field amplitudes were sampled at two representative locations: the geometric center of the VMR and an edge position near the tissue-facing sidewall. The field values were normalized to the corresponding H- and E-field amplitudes of the 10-ring VMR at the same sampling location. Additional line-profile simulations were performed along the axial direction through the coaxial cable cores to examine electric-field localization inside and outside the cable structure.

For scalability analysis, resonant rings with different diameters and gap configurations were simulated to quantify the relationship between resonance frequency and ring diameter. Both fixed single-gap and variable-gap configurations were evaluated. The variable-gap design was used to maintain comparable segment length and effective electromagnetic response as the ring diameter increased.

**Bench electromagnetic characterization**

Bench measurements were performed using a vector network analyzer (VNA, P5020B, Keysight Inc.). Reflection spectra S11 were measured using an inductive loop antenna placed about 50 mm above the VMR. For self-detuning characterization, the VNA excitation power was swept from -30 dBm to 10 dBm, and the resonance frequency shift was extracted from the measured S11 spectra. Transmission measurements S12 were performed to characterize the axial modal response of the VMR. One loop antenna was fixed on the bottom of the VMR, while the second antenna was translated along the central axis of the structure over a range of ~114 mm. At each axial position, the S12 transmission spectrum was recorded, and the resonance minima corresponding to the collective modes were extracted.

## MRI phantom experiments

MRI experiments were performed on a clinical 3.0 T MRI system (Philips Healthcare). Three agarose gel phantoms were prepared to match the three VMR sizes. The phantoms contained 1 wt % agarose gel and had diameters of 90 mm, 120 mm, and 150 mm, respectively. Three receiver configurations were compared: the VMR-augmented spine coil (VMR-SC), the spine coil alone, and a commercial 16-channel breast coil. For the VMR-SC and spine-coil-only measurements, each phantom was positioned at the same location relative to the spine coil to ensure identical loading and imaging geometry. For the breast coil measurements, each phantom was placed within one chamber of the breast coil housing, and the movable baffle was adjusted to provide the closest geometric fit for each phantom diameter. Each phantom was imaged in axial, sagittal, and coronal planes. Unless otherwise stated, MRI scans were acquired using a gradient-echo sequence. Detailed sequence parameters are provided in **Supplementary Table S4**.

## Parallel imaging experiments

Parallel imaging experiments were performed on the same 3.0 T MRI system using SENSE reconstruction. Acceleration factors of 2, 3, and 4 were evaluated. For each acceleration factor, images were acquired using the spine coil alone, the commercial 16-channel breast coil, and the VMR-SC configuration under otherwise identical imaging conditions. Representative images and quantitative SNR profiles were compared across the three receiver configurations. The dashed rectangular region in the images indicates the effective coverage region enclosed by the VMR. Detailed SENSE acquisition parameters are provided in **Supplementary Table S5**.

## SNR and uniformity analysis

SNR maps were calculated from magnitude images using a noise-normalized method. The signal image was divided by the standard deviation of the background noise measured from a signal-free region outside the phantom. For each imaging plane, the same noise ROI was used across receiver configurations when applicable. SNR profiles were extracted along the central horizontal and vertical lines of the selected ROI. Enhancement ratios were calculated as $SNR_{VMR\text{-}SC}/SNR_{SP}$ and $SNR_{VMR\text{-}SC}/SNR_{BR}$, where SP denotes the spine-coil-only configuration and BR denotes the commercial breast-coil-only configuration.

Spatial uniformity was quantified using the coefficient of variation, defined as the standard deviation divided by the mean SNR within the selected ROI. Difference maps between the VMR-SC and breast coil were calculated as $\Delta SNR$ = $SNR_{VMR\text{-}SC}$- $SNR_{BR}$. Positive $\Delta SNR$ regions indicate locations where the VMR-SC provided higher SNR than the commercial breast coil. The percentage of positive $\Delta SNR$ pixels were calculated within both the whole-phantom ROI and the VMR-covered ROI when applicable. All image post-processing and quantitative analyses were performed in MATLAB R2021b.

**B1+ compatibility and RF safety evaluation**

Transmit-field compatibility was evaluated using two complementary MRI experiments. First, flip-angle-dependent imaging was performed on a homogeneous phantom using flip angles ranging from 10° to 90° for the VMR-SC and Birdcage coil alone configurations. The mean signal intensity or SNR within a homogeneous ROI was extracted as a function of flip angle. Second, spatially resolved B1+ maps were acquired using a dual-TR method to assess whether the presence of the VMR introduced transmit-field distortion. The imaging parameters for flip-angle and B1+ mapping experiments are listed in **Supplementary Table S6**.

SAR simulations were performed in CST Microwave Studio using the time-domain solver and a modified full-body human voxel model. Local SAR averaged over 10 g of tissue was calculated using the MRI toolbox during post-processing and normalized to 1 W of accepted power. Simulations were performed with and without the VMR present under otherwise identical excitation conditions. The maximum SAR_10g values were compared with the IEC 60601-2-33 local SAR limit for normal operating mode.

Experimental heating validation was performed using MR thermometry based on the proton resonance frequency shift method. Phase images were acquired before and after high-power scanning sequences for the spine coil alone and VMR-SC configurations. Temperature-related phase evolution was compared across configurations to assess whether the VMR introduced measurable localized heating. Detailed thermometry sequence parameters and ROI definitions are provided in **Supplementary Fig. S11**.

**Ex vivo heterogeneous sample imaging**

A tomato sample was imaged using the commercial 16-channel breast coil alone and the VMR-SC configuration under otherwise identical conditions. T1-weighted turbo spin echo (T1wTSE) and T2-weighted turbo spin echo (T2wTSE) sequences were acquired on the same 3.0 T MRI system. Sequence parameters, including TR, TE, field of view, matrix size, slice thickness and number of averages, are provided in **Supplementary Table S2**. Images were compared qualitatively for signal intensity and visibility of internal structural features.

**Figure preparation**

Three-dimensional models and device renderings were generated using SolidWorks 2021 and KeyShot 2024. Scientific illustrations, annotations, and figure layouts were prepared using Adobe Illustrator 2025. Data plots, SNR maps, difference maps, and image profiles were generated using MATLAB R2021b. Electromagnetic field maps and surface-current distributions were exported from CST Microwave Studio Suite 2021.

**Data availability**

The data that support the findings of this study are available from the corresponding author upon reasonable request.

**Acknowledgements**

This research was supported by the Rajen Kilachand Fund for Integrated Life Science and Engineering. The authors thank Andrew Ellison for valuable assistance and discussions during MRI experiments. The authors thank the Boston University Photonics Center for technical support.

**Author contribution**

Y. Liu, X. Zhu, S. W. Anderson, and X. Zhang conceived the study. Y. Liu, X. Zhu, and X. Zhang designed and constructed the metamaterial. Y. Liu, X. Zhu and X. Zhang designed and conducted the bench measurements. Y. Liu, X. Zhu, C. LeBedis, S. W. Anderson, and X. Zhang conducted the MRI scans. All authors participated in discussing the results. Y. Liu, X. Zhu, S. W. Anderson, and X. Zhang drafted the manuscript. X. Zhang and S. W. Anderson provided project supervision.

**Competing interests**

The authors have filed patent applications on the work described herein, application No.: 16/002,458, 16/443,126, and 17/065,812. Applicant: Trustees of Boston University. Inventors: Xin Zhang, Stephan Anderson, Guangwu Duan, and Xiaoguang Zhao. Status: Active.

**Additional information**

Additional information is available for this work.